\documentclass[preprint,12pt,authoryear]{elsarticle}
\usepackage{epsfig}
\usepackage{amssymb}
\usepackage{amsmath}
\usepackage{url}
\usepackage{lineno}

\journal{arXiv}

\begin{document}

\begin{frontmatter}
\title{Morphology-based retrieval of local wave parameters and phase shifts in satellite-observed atmospheric disturbances} 

\author[sri]{Yuriy Klymenko\corref{cor1}}
\cortext[cor1]{Corresponding author}
\ead{yurklym@gmail.com}

\author[sri]{Alla Fedorenko}
\ead{fedorenkoak@gmail.com}

\author[sri]{Evgen Kryuchkov}
\ead{kryuchkov.ye@gmail.com}

\author[sri]{Oleg Cheremnykh}
\ead{oleg.cheremnykh@gmail.com}

\address[sri]{Space Research Institute of the National Academy of Sciences of Ukraine and the State Space Agency of Ukraine, 40 Acad. Glushkov Ave., Kyiv, 03187, Ukraine}

\begin{abstract}

Currently available methods for processing atmospheric wave disturbances are often based on integral wavelet transform techniques, which efficiently estimate spectral data but are fundamentally limited by the uncertainty principle. To address this limitation, this work proposes a novel morphological approach based on partitioning spatial disturbance profiles into separate, well-defined wave segments. By focusing exclusively on the local waveform morphology of a central half-wave and its two adjacent out-of-phase counterparts, and taking the modified Morlet wavelet as a primary example of a model wave packet, we successfully reconstruct the complete set of local wave parameters directly from the amplitudes and positions of these three dominant half-waves. Consequently, the theory demonstrates good agreement with observations, confirming the high accuracy of the proposed waveform morphology approach in reconstructing the localized structure of atmospheric disturbances. When applied to evaluating the phase shift between different wave field components, the method reveals that this shift is a weighted combination of the classical monochromatic phase difference and a morphological one determined by the internal symmetry of the underlying wave packets. Although the classical factor yields the dominant contribution, the symmetry-related factor provides a moderate correction to it. The applicability and accuracy of the approach are verified using observational data gathered over several orbits of the low-Earth-orbit satellite Dynamic Explorer 2 as it passed over Earth's polar caps. The obtained results confirm that the observed disturbances correspond to medium-scale acoustic-gravity waves propagating at a small angle to the horizontal plane, demonstrating the broader methodological potential of the proposed technique for analyzing a wide range of wave phenomena.
\end{abstract}




\end{frontmatter}


\section{Introduction and concept of the proposed approach}

Acoustic-gravity waves (AGWs) are one of the key mechanisms for the transfer of energy and momentum in the Earth's atmosphere (Hines, 1960; Nappo, 2002). AGWs make a significant contribution to the occurrence of atmospheric convection and turbulence, profoundly affecting the formation of weather systems and other atmospheric processes (Medvedev and Yi\u{g}it, 2019; Liu et al., 2024). Their sources can include both influences from ``below'' (such as tropical convection, cyclonic and thunderstorm activity, as well as seismic and volcanic events) and influences from ``above'' (auroral currents, Joule heating of the ionosphere, particle precipitation, solar terminator, magnetospheric disturbances, and others) (Vadas et al., 2023; Achatz et al., 2024; Cheremnykh et al., 2023; Cheremnykh et al., 2024). These waves manifest as fluctuations in atmospheric thermodynamic parameters (such as density, temperature, and pressure) and variations in kinematic properties, specifically the velocity components for an air parcel or its neutral and ionized constituents individually. All such physical disturbances belonging to the same wave field will hereafter be referred to as the wave field components. These components propagate across a wide range of wavelengths, spanning from hundreds to several thousand kilometers (Hines, 1960; Nappo, 2002).

Low-orbit \textit{in situ} satellite measurements play a crucial role in AGW diagnostics. Unlike ground-based observations (optical, radiophysical, and radar) -- which provide continuous temporal dynamics at a fixed location or local vertical profiles (e.g., altitude profiles of plasma density and velocity obtained via ionosondes and incoherent scatter radars) -- satellite data yield a spatial profile of atmospheric disturbances along the spacecraft's orbit trajectory (Johnson et al., 1995; Innis and Conde, 2002). Because the orbital velocity of a low-Earth orbit satellite ($V_{sat} \sim 7.8\text{ km/s}$ at an altitude of 250--500 km) significantly exceeds the phase speed of typical AGWs ($v_{ph} \ll V_{sat}$), the measured time series can be reliably interpreted as a quasi-instantaneous 1D spatial snapshot of the wave field. Nevertheless, potential temporal distortion of this spatial profile must be accounted for, as it depends on whether the AGW propagates along or opposite to the satellite's motion path.

Once the large-scale background trend of the undisturbed atmosphere is removed, the spatial wave profile (such as temperature or density fluctuations) recorded by satellite instruments takes the form of quasi-periodic oscillations localized within a specific region of space. Most traditional methods for acoustic-gravity wave parameter estimation and retrieval rely on plane monochromatic wave approximations or local wave analysis (Sakib and  Yi\u{g}it, 2022; Li et al., 2022). Within the local analysis framework, an observed profile is segmented into short intervals, over each of which the signal is approximated as a monochromatic wave with a constant frequency and wavevector. This approach has proven highly effective for addressing fundamental problems in atmospheric physics: it has enabled the construction of global wave activity maps, the calculation of vertical AGW momentum and energy fluxes in the mesosphere and thermosphere, and the assessment of wave driving in upper-atmosphere circulation (Sakib and Yi\u{g}it, 2022; Li et al., 2022).

In contrast to the local approach, representing real disturbances from localized sources (e.g., a weather front, a solar terminator, or a lunar shadow) as wave packets is physically more sound. The wave packet concept has demonstrated its efficiency in solving direct problems: modeling the ionospheric response to the passage of the solar terminator (Afraimovich et al., 2010), as well as the generation and dispersion of AGW wave trains induced by the movement of the lunar shadow during solar eclipses (Schoennagel et al., 2006). Crucially, such studies deal specifically with the \textit{direct problem} -- calculating the spatio-temporal evolution of a predefined wave packet.

To analyze the structure of such packets from experimental profiles, wavelet analysis or windowed Fourier transform methods are frequently applied (e.g., \v{S}auli et al., 2007). These approaches efficiently estimate the local spectral composition, which is determined from the positions and inherent spectral broadening of local maxima in the two-dimensional continuous wavelet transform spectrogram. However, due to the uncertainty principle, such spectral estimates face a fundamental limitation: enhancing spectral resolution inevitably degrades spatial localization. Recent studies (He et al., 2024; Wright et al., 2024) specifically detail the specific limitations of wavelet analysis when retrieving local parameters of non-stationary waves, emphasizing the inherent constraint between spatial and temporal  resolution, which is additionally complicated by background wind shear and the exponential growth of AGW amplitude with altitude.

Consequently, while traditional wave parameter retrieval techniques are well-developed for monochromatic waves and local spectral analysis (Sakib and Yi\u{g}it, 2022; Li et al., 2022), their application to wave packets remains restricted either to solving direct problems or to qualitative wavelet visualization. As for the \textit{inverse problem} -- namely, the accurate reconstruction of the full spectral composition, propagation direction, and phase structure of the wave packet directly from its spatial profile -- it has previously been formulated and successfully solved exclusively within the plane monochromatic wave approximation (Johnson et al., 1995; Innis and Conde, 2002; Fedorenko and Kryuchkov, 2014; Klymenko et al., 2021). At the same time, for a more realistic wave model -- a compact non-monochromatic wave packet -- the inverse problem of retrieving the original spectrum has hardly been addressed in the current literature.

To fully characterize the wave packet it is essential to analyze multiple wave field components simultaneously rather than focusing on a single component. When processing such multi-component observations within traditional frameworks, individual spectrograms are first compared to verify a shared spectral composition. Once confirmed, subsequent cross-component analysis is routinely performed using Cross-Wavelet Transform and Wavelet Coherence techniques. These methods provide scale-dependent estimates of phase shifts and mutual coherence between the wave field components as a function of spatial scale and position, inherently averaging these characteristics over a finite spatial domain.

In this paper, we propose a new analytical approach for processing satellite observational data of wave field components. Operating independently of conventional numerical wavelet schemes, our method analytically reconstructs the local spectral parameters of the signals and directly resolves the phase shifts between them. The core of this approach lies in partitioning the continuous wave profile into distinct, individual wave segments, each consisting of several half-waves. By shifting the focus of the analysis from integral transforms to the waveform morphology of these localized elements, the method effectively bypasses the constraints of the uncertainty principle, achieving maximum spatial resolution within these narrow segments.

While a traditional wavelet spectrogram identifies the wide localization region of wave activity -- providing a smoothed representation of its spectral distribution in this region -- our waveform morphology approach substantially narrows down the spatial domain, capturing the localized wave packet structure and extracting its local spectrum. Whereas the cross-wavelet spectrogram reveals broad, fuzzy zones of coherence, our approach validates this coherence specifically within this localized segment and directly yields the phase shift between the field components. This analytical and morphological framework offers a mathematically transparent way to interpret experimental data, yielding exact local spectral parameters and providing a clear insight into relative phase shifts.

For illustration, Figure 1 shows a representative example of partitioning wave profiles into individual wave segments. These profiles, corresponding to fluctuations in temperature and density as well as the vertical velocity component, were recorded on orbit 8132 of the low-Earth orbit Dynamic Explorer 2 (DE2) satellite during its crossing of the Antarctic (representing a portion of the total 1.5-hour orbital period). The waveform partitioning algorithm begins by identifying the global extremum within a selected region of each profile, which is interpreted as the main half-wave of the first localized wave packet. Next, the structure of the profile on either side of this dominant half-wave is evaluated, and the equidistance of the two accompanying side half-waves of opposite sign is visually verified. This verification is based on the rigorous quantitative symmetry criteria formulated later in Section 6. The spatial coordinates and amplitudes of these three key half-waves are then recorded, and this extraction procedure is repeated cyclically over the remaining unsegmented intervals of the profile.

\begin{figure}[!htbp]
\centering
\includegraphics[width=1.0\columnwidth]{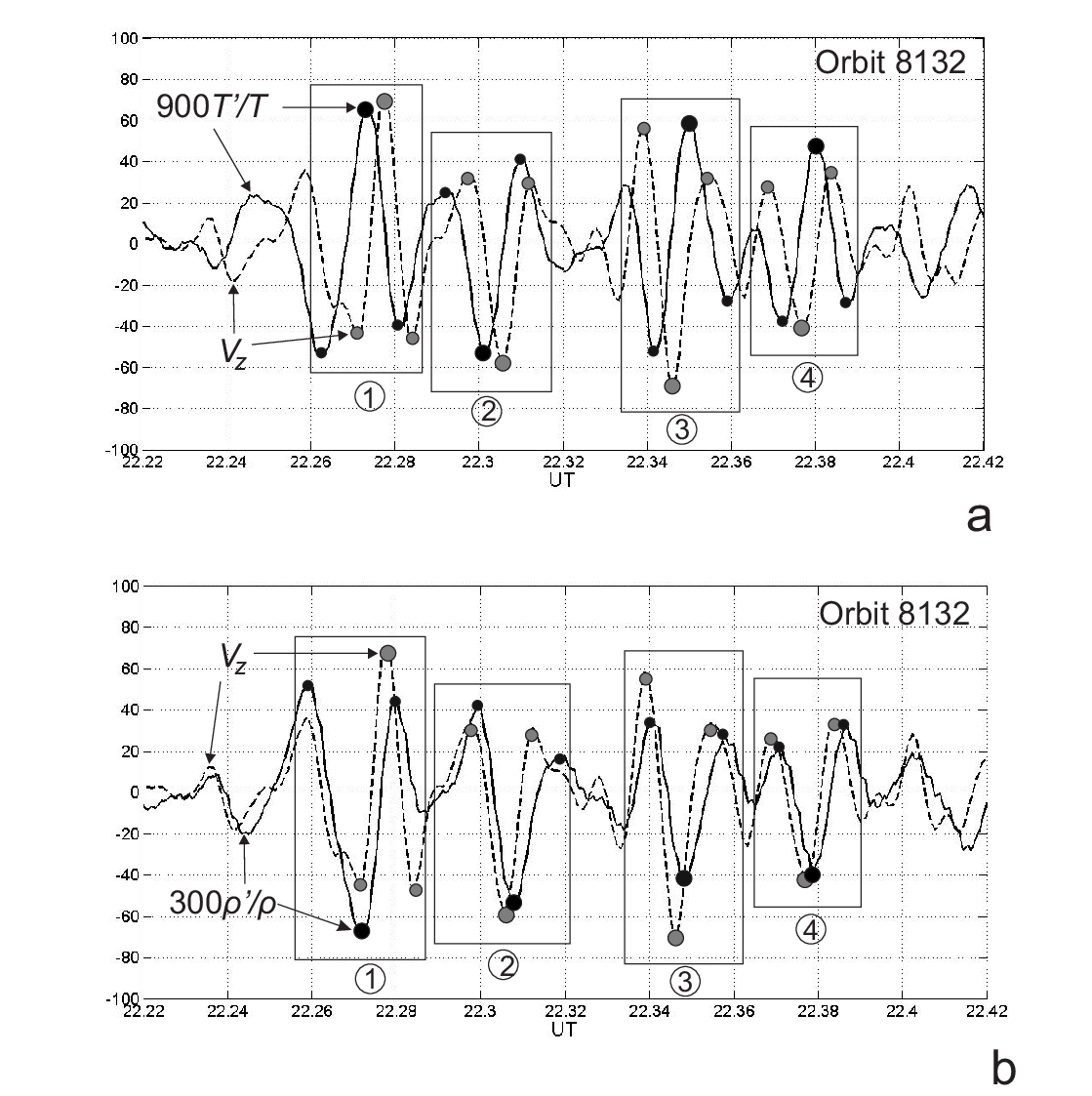}
\caption{Example of the decomposition of the spatial wave profiles recorded on orbit 8132 of the DE2 satellite (southern polar cap crossing) into four isolated wave packets (outlined by rectangles and numbered): (\textbf{a}) vertical velocity and relative temperature fluctuations; (\textbf{b}) vertical velocity and relative density fluctuations along the same trajectory segment. Large black and gray circles mark the central dominant extrema of the respective fields within each packet. Smaller circles indicate the positions of the adjacent side half-wave extrema, which are nearly equidistant from the central extremum and opposite in sign. Universal Time (UT) along the horizontal axis corresponds to the spatial coordinate along the satellite's trajectory when multiplied by the spacecraft's velocity $V_{sat}$.}
\end{figure}

Spatially complex or highly distorted segments should be excluded from our analysis. In the specific case of Figure 1, these degraded regions are located at the outer boundaries of the entire interval. However, in actual observational data, such distorted segments can occur at any region of the measured profile. Within these complex intervals, wave packets may overlap, while the signal may be heavily affected by localized noise or wave processes of a different physical nature. This physical ambiguity limits the direct application of our method within these intervals, making preliminary filtering or alternative processing techniques necessary.

In Figure 1, we successfully identified four distinct segments numbered 1 through 4. They correspond to physical wave packets, which cross the satellite's trajectory and are recorded in situ by the onboard instruments. In Sections 2-4, we formulate the wave packet framework for such segments. Although this formulation ultimately yields the standard analytical representation of a classical quasi-monochromatic wave packet, our approach offers a deeper physical interpretation of the localized wave event. Specifically, instead of relying on abstract spectral representations, it provides a direct, self-consistent link between the observed spatial profile of a single packet and the underlying characteristics of the interfering waves that form its envelope.

In Section 5, we reformulate the wave packet description in terms of a wavelet-like representation by shifting the spatial coordinate origin to the specific packet's center. Crucially, this spatial translation alters the wave phase, transforming the original phase into a wavelet one. The wavelet phase directly reflects the internal symmetry of the wavelet -- specifically, its dominant half-waves -- and is hereafter referred to as the geometric phase. This representation is highly convenient for the analysis of the spatial profile of the wave field components, which are analyzed independently of each other. However, to find the physical phase shifts between these wave field components, a transition back to the initial reference phase is required.

At the next step, we approximate the shape of the slowly varying envelope by the widely used Gaussian function, which is the most widely accepted form for describing the envelope of real atmospheric wave packets. This choice establishes a direct analytical bridge to the modified Morlet wavelet. This approximation enables an analytical reconstruction of the key wavelet parameters in Section 6. The entire procedure requires only the positions and amplitudes of the three dominant half-waves located within the local regions of the observed disturbances. Crucially, two of these parameters -- the wavenumber and the spectral bandwidth parameter -- characterize the underlying wave spectrum; therefore, they must remain identical for all analyzed components, provided these components originate from the same wave event. The remaining three parameters -- the spatial center of the envelope, its peak amplitude, and the geometric phase -- are determined for each component individually. Importantly, our framework is not restricted to a Gaussian profile; this function can be readily replaced by by any physically motivated bell-shaped envelope.

In Section 7, we present a comparison between the satellite-observed profiles and the reconstructed analytical waveforms. Specifically, for each isolated segment, we retrieve the wave packet parameters and substitute them into the theoretical formulas to reconstruct the local wave profile. These local contributions are then summed across all segments to yield the total reconstructed profile, which is ultimately compared against the initial observational data. This validation is performed using observational data from three distinct orbits of the DE2 satellite. The results demonstrate that the analytical curves are in excellent agreement with the measured data.

In particular, when applied to the complex disturbance patterns shown in Figure 1, this approach reveals underlying regularities and brings order to the apparent chaos of wave processes. The effectiveness of the proposed morphology-based theory is clearly demonstrated in Figure 4 below, which compares the theoretical profiles with satellite observation data and illustrates the envelopes of the isolated wave packets constructed from the theory. As can be seen, the analytical model captures the observational data with remarkable accuracy, faithfully reproducing the observed oscillatory behavior within all theoretically derived envelopes, thereby confirming the capability to reconstruct the spatial structure of wave fields across the entire analyzed region.

Up to this point, the profiles of different wave field components have been analyzed independently of one another based on their wavelet description. In Section 8, we investigate the physical phase shift between these wave field components, which requires returning to the initial phases of the original wave packets. We establish that this phase shift is a weighted combination of two distinct contributions: the classical monochromatic phase difference (defined as $2\pi $ times the spatial distance between the respective central extrema divided by wavelength) and the relative shift between the local geometric phases that govern the symmetry of each wavelet. This symmetry-related contribution vanishes if the profiles are identical in shape. The classical term contributes the dominant share (with its weight approaching unity in the monochromatic limit), while the geometric term provides a moderate yet noticeable correction of up to 10--15 degrees.

By accounting for this geometric correction, we can successfully extract the true, undistorted physical phase shifts from the real satellite observations. Knowing these phase shifts is essential for accurate wave analysis. In particular, when applied to satellite measurements where only a one-dimensional cross-section of the wave field is available, these true physical phase shifts serve as key input parameters for our theoretical models. To illustrate this application, we focus on the following case.

For monochromatic AGWs freely propagating in a stratified, non-dissipative hydrodynamic medium within the framework of Hines' model, we demonstrated in a previous study (Klymenko et al., 2021) that a complete reconstruction of their spectral parameters (horizontal and vertical wavelengths, as well as intrinsic frequency) can be achieved using only two phase shifts -- specifically, between the vertical velocity and the relative fluctuations of either temperature or density. Crucially, these physical phase shifts are independent of the choice of coordinate system, whether spacecraft-centered or Earth-fixed. Based on this invariance, we introduced a phase diagnostic diagram. Plotted in the Cartesian plane of these two phase shifts, this diagram visually identifies both the AGW type (acoustic or gravity) and its vertical propagation direction (upward or downward) once the corresponding phase point is mapped onto it. The remaining spectral parameters are subsequently reconstructed analytically. Since this approach is essential for our analysis, we present its main steps in Section 9.

In Section 10, we determine all AGW spectral parameters using data from the three DE2 satellite orbits. All these orbits correspond to the satellite passes over the Earth's polar caps. For the aforementioned orbit 8132, the parameters are obtained directly by substituting the calculated phase shifts into the initial analytical formulas. For the remaining two orbits, we significantly simplify these formulas based on the distinct pattern of the observed phase shifts. All these calculations confirm that the detected disturbances are medium-scale gravity waves traveling nearly horizontally.

In  Section 11, we summarize the main results of this study and provide a brief discussion of their implications.

\section{Monochromatic solutions of linearized hydrodynamic equations for AGWs}

For an isothermal stratified atmosphere with a constant chemical composition, no wind, and no dissipation, the motion of an infinitesimal volume is described by linearized hydrodynamic equations (Klymenko et al., 2021):
\begin{subequations}
\label{eq:main}
\begin{align}
    \label{eq:sub_a}
     &\frac{\partial V_{z} }{\partial t} +gH\frac{\partial }{\partial z} \left(\frac{\rho '}{\rho } +\frac{T'}{T} \right)-g\frac{T'}{T} =0, \\
    \label{eq:sub_b}
    &\frac{\partial V_{x} }{\partial t} +gH\frac{\partial }{\partial x} \left(\frac{\rho '}{\rho } +\frac{T'}{T} \right)=0, \\
    \label{eq:sub_c}
    &\frac{\partial }{\partial t} \left(\frac{\rho '}{\rho } \right)+\mathrm{div}\mathbf{V}-\frac{V_{z} }{H} =0, \\
    \label{eq:sub_d}
    &\frac{\partial }{\partial t} \left(\frac{T'}{T} \right)+\left(\gamma -1\right)\mathrm{div}\mathbf{V}=0.
\end{align}
\end{subequations}
Here, $V_{z} $ and $V_{x} $ are the vertical and horizontal velocity components (the $z$-axis is directed vertically upward, and the $x$-axis is oriented along the horizontal motion of the atmospheric particles), $g$ is the acceleration due to gravity, and $H=RT/m_{0} g$ is the atmosphere scale height (equal approximately to 50 km under average atmospheric conditions for satellite altitudes of 250--500 km). Additionally, $\rho $($\rho '$) and $T$($T'$) are the background (perturbed) density and temperature, respectively, $R$ is the universal gas constant, $m_{0} $ is the mean molar mass of the atmosphere, and $\gamma $ is the adiabatic index. Here, we set $\gamma =5/3$, corresponding to a monatomic gas approximation for the upper atmosphere at the satellite altitudes of interest, where photodissociation leads to the dominance of atomic oxygen.

If the acoustic-gravity wave is freely propagating and monochromatic, the solutions to equations (1a)--(1d) are sought in the form of the traveling waves
\begin{equation}\label{2}
\frac{V_{z} }{M^{V_{z} } } =\frac{V_{x} }{M^{V_{x} } } =\frac{\rho '/\rho }{M^{\rho '/\rho } } =\frac{T'/T}{M^{T'/T} } =Ae^{z/2H} e^{i(\omega t-\mathbf{k}\mathbf{r})}.
\end{equation}
Here, $M^{\mu } $ ($\mu =V_{z} ,\, V_{x} ,\, \rho '/\rho ,\, \, T'/T$) are the currently unknown polarization coefficients, $A$ is an arbitrary complex constant common to all components of the wave field, $\mathbf{k}=(k_{z} ,k_{x} )$ is the wave vector with its vertical and horizontal components, $\mathbf{r}=(z,x)$ is the radius vector, and $\omega $ is the angular frequency of the wave.

Substituting (2) into differential equations (1a)--(1d) yields a system of linear homogeneous  equations with respect to unknowns $M^{\mu } $. The condition for a non-zero solvability of this system is that its determinant equals zero. This leads to the dispersion relation
\begin{equation}\label{3}
\left(\frac{\omega }{N} \right)^{4} -\frac{\gamma ^{2} }{\gamma -1} \left(\frac{1}{4} +\left(k_{z} H\right)^{2} +\left(k_{x} H\right)^{2} \right)\left(\frac{\omega }{N} \right)^{2} +\frac{\gamma ^{2} }{\gamma -1} \left(k_{x} H\right)^{2} =0,
\end{equation}
which implicitly determines the dependence of the frequency on the wave vector components: $\omega =\omega (k_{z} ,k_{x} )$. In Eq. (3), the value $N=c_{s} \sqrt{\gamma -1} /H\gamma $ is the Brunt--V\"{a}is\"{a}l\"{a} frequency, and $c_{s} $ is the sound speed in the atmosphere. For satellite altitudes: $c_{s} \approx 800\div 900$m/s and $N \approx 0.01$s$^{-1}$ under average atmospheric conditions.

Solving Eq. (3) with respect to $\omega $, we obtain the dispersion relation for the acoustic (+) and gravity (-) branches of AGWs:
\begin{equation}\label{4}
\begin{aligned}
\left(\frac{\omega}{N}\right)_{\pm}^{2} &= \frac{\gamma^{2}}{2(\gamma - 1)} \Big[\frac{1}{4} + (k_{x} H)^{2} + (k_{z} H)^{2} \\
&\quad \pm \sqrt{\left(\frac{1}{4} + (k_{x} H)^{2} + (k_{z} H)^{2}\right)^{2} - \frac{4(\gamma - 1)}{\gamma^{2}} (k_{x} H)^{2}} \Big]
\end{aligned}.
\end{equation}

If the dispersion relation holds, the polarization equations become linearly dependent and can be used to obtain the polarization coefficients (Hines, 1960; Klymenko et al., 2021):
\begin{subequations}
\label{eq:main}
\begin{align}
    \label{eq:sb_a}
     &M^{V_{z} } (\omega ,k_{z} ,k_{x} )=2H\omega \left(\omega ^{2} -c_{s}^{2} k_{x}^{2} \right), \\
    \label{eq:sb_b}
    &M^{V_{x} } (\omega ,\, k_{z} ,k_{x} )=2Hc_{s}^{2} k_{x} \omega \left[k_{z} -\frac{i(2-\gamma )}{2\gamma H} \right], \\
    \label{eq:sb_c}
    &M^{\rho '/\rho } (\omega ,k_{z} ,k_{x} )=2\omega ^{2} (k_{z} H)-i\left[\omega ^{2} -2\left(1-\gamma ^{-1} \right)c_{s}^{2} k_{x}^{2} \right], \\
    \label{eq:sb_d}
    &M^{T'/T} (\omega ,k_{z} ,k_{x} )=(\gamma - 1)\left[2\omega ^{2} (k_{z} H)+i(\omega ^{2} -2\gamma ^{-1} c_{s}^{2} k_{x}^{2})\right].
\end{align}
\end{subequations}

The final substitution of (5) into (2) yields the solution
\begin{equation} \label{6}
\mu \left(\mathbf{r},t\right)=\left|A\right|e^{z/2H} M^{\mu } (\omega ,\mathbf{k})e^{i\left[\alpha +\omega (\mathbf{k})t-\mathbf{k}\mathbf{r}\right]}, \quad \mu =V_{z} ,\, V_{x} ,\, \rho '/\rho ,\, T'/T,
\end{equation}
for the wave field components of a plane monochromatic wave propagating in the atmosphere. Here, the complex amplitude $A$ is represented in exponential form: $A=|A|\exp(i\alpha)$.

As can be seen, the phase shifts between monochromatic solutions (6) are determined by the differences between the arguments of the complex functions (5a)--(5d). For convenience, we choose the first profile $V_{z} $ as a reference, since its corresponding polarization function is purely real. This allows us to find the phase shifts for the remaining monochromatic profiles directly from the equations
\begin{subequations}
\label{eq:main}
\begin{align}
    \label{eq:s_a}
     &\tan\Delta _{\rho ,V_{z} } =-\frac{\omega ^{2} -2\gamma ^{-1} (\gamma -1)c_{s}^{2} k_{x}^{2} }{2\omega ^{2} (k_{z} H)}, \\
    \label{eq:s_b}
    &\tan\Delta _{T,V_{z} } =\frac{\omega ^{2} -2\gamma ^{-1} c_{s}^{2} k_{x}^{2} }{2\omega ^{2} (k_{z} H)}, \\
    \label{eq:s_c}
    &\tan\Delta _{V_{x} ,V_{z} } =\frac{\gamma -2}{2\gamma (k_{z} H)}.
\end{align}
\end{subequations}

\section{AGW packet composed of monochromatic waves}

Let us introduce the unit vector $\mathbf{e}_{x'} $ indicating the direction of wave propagation. In this case, $\mathbf{k}\equiv k\mathbf{e}_{x'} $ and $\mathbf{k}\mathbf{r}=kx'$, where $k$ is the magnitude of the wave vector (hereinafter referred to as the wavenumber), and $x'$ is the longitudinal spatial coordinate measured along the direction of wave propagation. Next, we rewrite the dispersion relation as an explicit function of the wavenumber: $\omega \equiv \omega (\mathbf{k})=\omega (k\mathbf{e}_{x} )\to \omega (k)$. All this allows us to transform expression (6) into the following explicit dependencies:
\begin{equation}\label{8}
\mu \left(x',z,t\right)=\left|A\right|e^{z/2H} M^{\mu } [\omega (k),k]e^{i\left[\alpha +\omega (k)t-kx'\right]}.
\end{equation}

Expression (8) represents a particular monochromatic solution. According to the superposition principle, a more general solution can be constructed as a finite sum (linear combination) of such monochromatic waves.

To describe a wave packet, let us now consider a large number of these waves with the same dispersion propagating along the unit vector $\mathbf{e}_{x'}$. Suppose that their wavenumbers are uniformly distributed over a finite spectral interval $[k_0 - \Delta k, k_0 + \Delta k]$, where $k_{0}$ is the central (carrier) wavenumber and $\Delta k$ is its spectral half-width. By introducing the discretization step $\Delta k/\mathcal{N}$ and letting $\mathcal{N} \to \infty$, we can pass from discrete summation to integration and write
\begin{equation}\label{9}
\frac{\mu \left(x',z,t\right)}{\cal{N}} =\frac{e^{z/2H} }{\Delta k} \int _{k_{0} -\Delta k}^{k_{0} +\Delta k}\left|A(k)\right|M^{\mu } (k)e^{i[\alpha (k)+\omega (k)t-kx']} dk.
\end{equation}
In further analysis, the partial factor $\mathcal{N}$ will be omitted.

In the packet solution (9), the amplitude $A(k)$ and phase $\alpha(k)$ represent the spectral distributions of the waves constituting the packet. To keep the physical meaning of the wave packet clear, we assume the amplitude spectrum is bell-shaped. In other words, it reaches its maximum at the central wavenumber $k_{0} $ and decreases symmetrically on both sides. The phase spectrum $\alpha (\omega )$, on the other hand, we leave completely arbitrary.

From now on, we assume that $\Delta k$ is small compared to $k_{0} $. Let us rewrite the polarization factor in exponential form, $M^{\mu } \equiv \left|M^{\mu } \right|e^{i\arg (M^{\mu } )} $, and expand all the integrands from Eq. (9) in a Taylor series in powers of the small parameter $(k-k_{0}^{} )$. The primary contribution to the integral comes from the zeroth-order approximation of the amplitude functions: $\left|A(k)M^{\mu } (k)\right|\approx \left|A(k_{0} )M^{\mu } (k_{0} )\right|$. In the phase of the integrand, which equals $\Phi (k)=\alpha (k)+\arg M^{\mu } (k)+\omega (k)t-kx'$, we retain both the zeroth-order and linear terms: $\Phi (k)\approx \Phi (k_{0} )+\Phi '(k)\left|_{k=k_{0} } \right. (k-k_{0} )$. Here,
\[\begin{array}{l} {\Phi (k_{0} )=\alpha (k_{0} )+\arg M^{\mu } (k_{0} )+\omega (k_{0} )t-k_{0} x',} \\ {\Phi '(k)=\frac{\partial }{\partial k} \left[\alpha (k)+\arg M^{\mu } (k)+\omega (k)t\right]-x'.} \end{array}\]

Keeping in mind that the derivative $\left. \partial \omega /\partial k\right|_{k=k_{0} } $ represents the group velocity $V^{gr} $ of the packet, we finally arrive at the following expression:
\begin{equation}\label{10}
\mu \left(x',z,t\right)=e^{z/2H} \Omega ^{\mu } (x'\Delta k,t)\left|M^{\mu } (k_{0} )\right|e^{i\left[\alpha (k_{0} )+\omega (k_{0} )t-k_{0} x'\right]}.
\end{equation}
Here, the real function
\begin{equation}\label{11}
\Omega ^{\mu } (x'\Delta k,t)=\frac{1}{\Delta k} \int _{-\Delta k}^{\Delta k}\left|A(\kappa )\right|\cos \kappa \left(x'-V^{gr} t-x_{0}^{\mu } \right)d\kappa
\end{equation}
defines the spatial envelope of the wave packet (written in terms of the shifted integration variable $\kappa =k-k_{0} $, with its geometric center located at
\begin{equation}\label{12}
x_{0}^{\mu } =\left. \frac{\partial }{\partial k} \left[\alpha (k)+\arg M^{\mu } (k)\right]\right|_{k=k_{0}}.
\end{equation}

As can be seen from Eq. (11), the envelopes of all measured physical components have the same spatial form. This form is defined by the integral (11) and can be explicitly determined once the specific function under the integral sign is chosen. All wave packets propagate with the same group velocity $V^{gr} $. In the limit $\Delta k\to 0$, the envelope (11) becomes constant, and expression (10) reduces to the plane monochromatic wave (8) at the carrier wavenumber $k_{0} $.

Now we focus on the spatial position of the envelope center given by formula (12). The first term yields a constant shift common to all components of the wave field. The second, polarization term in (12), however, causes the centers of different envelopes to be shifted relative to each other. Consequently, the envelopes for two wave field components will coincide only if the arguments of their respective polarization functions have the same derivative at the carrier wavenumber $k_{0} $. Since this condition is rarely met, the envelope centers differ for each wave field component. This fact is accounted for by introducing the upper index in the envelope notation.

Finally, expressions (10) and (8) demonstrate complete phase coincidence. This implies that the phase shifts in the packet representation are identical to those in the monochromatic description.

\section{Adaptation of packet profiles to satellite observations}

The wave packet (10) with the envelope (11) defines the spatiotemporal profile of the disturbance. To trace these disturbances along the satellite's trajectory (defined by the unit vector $\mathbf{e}_{X} $ and longitudinal coordinate $X$), one would generally project the the wave parameters $k_{0} $ and $\Delta k$ onto the spacecraft's orbit using a certain angle between orts $\mathbf{e}_{X} $ and $\mathbf{e}_{x'} $. However, introducing new symbols for these projections would make the formulas too complicated and difficult to read. For simplicity, we assume that $k_{0} $ and $\Delta k$ in the formulas above already represent their projections onto the satellite's path. This convention will not cause any confusion.

Before writing the necessary expression for the wave packet projection onto the satellite's orbit, we must transform it from complex to real form, since the measured physical parameters are always real. To do this, we take a linear combination of the real and imaginary parts of (10) with coefficients, which are identical for all wave components. Due to the linearity of the hydrodynamic equations (1a)-(1d), such a combination remains a valid solution.

Finally, we fix both the altitude of the satellite orbit and the signal registration time. In this way, we obtain the desired one-dimensional spatial dependence with respect to the longitudinal coordinate $X$ of the satellite:
\begin{equation}\label{13}
\begin{split}
\mu(X) &= C^{\mu} \Omega^{\mu} (X\Delta k)\cos\left(k_{0} X+\varphi^{\mu}\right), \\
\Omega^{\mu}(X\Delta k) &= \frac{1}{\Delta k} \int_{-\Delta k}^{\Delta k}\left|A(\kappa)\right|\cos \kappa\left(X-X_{0}^{\mu}\right)d\kappa .
\end{split}
\end{equation}
Specifically, this expression must be directly compared with the corresponding wave field component observed by the satellite. In this expression, $C^{\mu } $ and $\varphi ^{\mu } $ are the amplitude and initial phase.

Now let us return to the phases of the disturbances, or rather, to the phase differences between them. As shown at the end of the previous section, the transition from a monochromatic solution to a wave packet does not change the phase differences between the components. Similarly, projecting the disturbances onto the satellite's orbit does not alter this phase difference either. This is clearly illustrated in Figure 2. In fact, this implies that the phase differences of the physical components measured by the satellite will be identical to those of the monochromatic waves, for which the description in Section 2 holds true.
\begin{figure}[htbp]
\centering
\includegraphics[width=1.0\columnwidth]{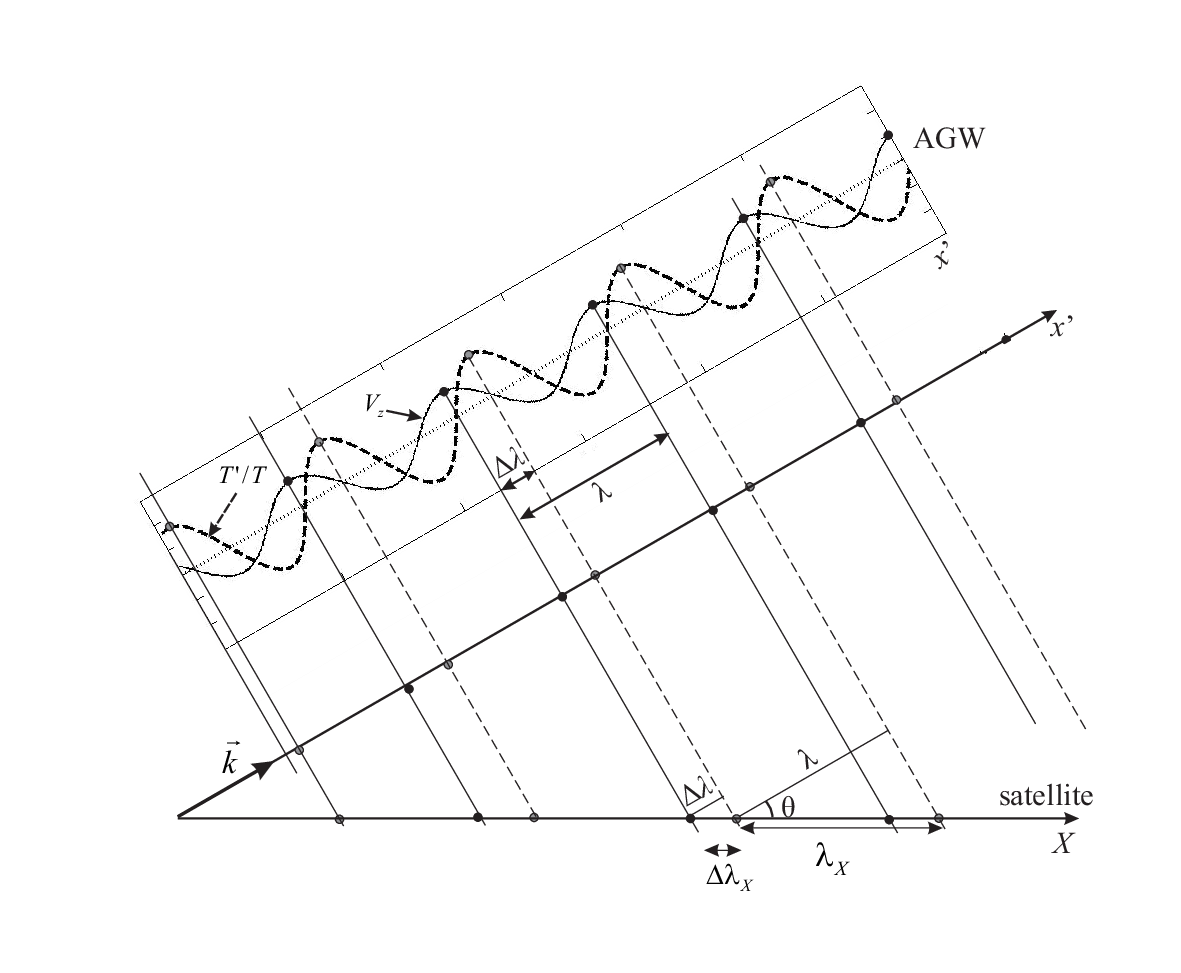}
\caption{Projections of two wave field disturbances of freely propagating AGW onto the satellite orbit. A satellite passing along its orbit will register the exact same phase shifts between the recorded profiles as in the original wave structure: $\left. \Delta \varphi \right|_{AGW} =-2\pi \frac{\Delta \lambda }{\lambda } =-2\pi \frac{\Delta \lambda _{X}^{} }{\lambda _{X}^{} } =\left. \Delta \varphi \right|_{sat} $.}
\end{figure}

This important statement allows us to view equations (7a)-(7c) from another perspective. The invariance of the phase differences makes it possible to determine all the left-hand sides of these formulas from the satellite observations. Treating them as known constants, the three equations (7) transform into a linear system with respect to the two dimensionless parameters $k_{z} H$ and\textbf{ }$(c_{s} k_{x} /\omega )^{2} $\textbf{.} The overdetermined nature of this system (three equations for two unknowns) implies that
\[\left(\gamma -1\right)\, \tan\Delta _{T,V_{z} } +\tan\Delta _{\rho ,V_{z} } -\gamma \tan\Delta _{V_{x} ,V_{z} } =0. \]
Notably, this formula includes only the adiabatic constant $\gamma $ and no other physical characteristic of the atmosphere.

\section{Wavelet form of the wave packet and its concretization}

Equation (13) represents a spatial packet with an explicitly defined envelope shape. We rewrite it in the relative coordinate $x^{\mu } =X-X_{0}^{\mu } $, as is commonly done in wavelet analysis. Then,
\begin{equation}\label{14}
\begin{split}
\Psi^{\mu} \left(x^{\mu}\right) &= C^{\mu} \Omega (x^{\mu} \Delta k)\cos (k_{0} x^{\mu} +\delta^{\mu}), \\
\Omega (x^{\mu} \Delta k) &= \frac{1}{\Delta k} \int_{-\Delta k}^{\Delta k}\left|A(\kappa)\right|\cos (\kappa x^{\mu})d\kappa.
\end{split}
\end{equation}
We introduce here the new symbol $\Psi$ (standard in wavelet notation) and define the new phase:
\begin{equation}\label{15}
\delta ^{\mu } =\varphi ^{\mu } +k_{0} X_{0}^{\mu },
\end{equation}
which determines the internal symmetry of the wavelet. This linear shift between the symmetry phase $\delta ^{\mu } $ and the initial phase $\varphi ^{\mu } $ will be highly useful in our further analysis.

For the subsequent analysis, we concretize the functional form of the envelope, although it could generally be any bell-shaped function. Let us assume that the amplitude spectrum has a Gaussian profile $\left|A(\kappa )\right|\sim \exp (-\kappa ^{2} /2\sigma ^{2} )$ with a standard deviation $\sigma $, which is assumed to be smaller than $\Delta k/4$. By evaluating the corresponding integral in Eq. (14), we obtain, with high accuracy:
\begin{equation} \label{16}
\Psi ^{\mu } \left(x^{\mu } \right)=C^{\mu } \exp{\left(-\frac{\sigma ^{2} x_{\mu }^{2} }{2} \right)} \cos (k_{0} x^{\mu } +\delta ^{\mu } ),\quad  x^{\mu } =X-X_{0}^{\mu }.
\end{equation}

 In this way, we arrive at a modified Morlet wavelet with five unknown parameters. Two of them -- the local carrier wavenumber $k_{0} $ and the spatial localization parameter $\sigma $ of the Gaussian envelope -- are spectral characteristics of the wavelet itself and therefore should manifest identically in all recorded wave field components. The other three parameters -- the disturbance amplitude $C^{\mu } $, the spatial position $X_{0}^{\mu } $ of the envelope, and the local geometric phase $\delta ^{\mu } $ determining its internal symmetry -- must be strictly individual and calculated separately for each measured wave field component.

\section{Parameterization of the wavelet via its three dominant half-waves}

Let us attempt to express all five wavelet parameters in (16) in terms of the amplitudes and positions of its three dominant half-waves: the global central half-wave with coordinate $x_{0}^{\mu } $ (located near the packet center $X_{0}^{\mu } $) and two adjacent side half-waves with coordinates $x_{-1}^{\mu } $ and $x_{+1}^{\mu } $. The amplitudes of these side half-waves must be smaller than that of the global one and have the opposite sign. For certainty, the case where the central half-wave is a global maximum and the two side half-waves correspond to troughs is shown in Figure 3. Conversely, the case where the primary half-wave is a trough and the side half-waves are local maxima is easily accounted for mathematically by adding a phase of $\pi $ to the symmetry phase $\delta ^{\mu } $.
\begin{figure}[htbp]
\centering
\includegraphics[width=1.0\columnwidth]{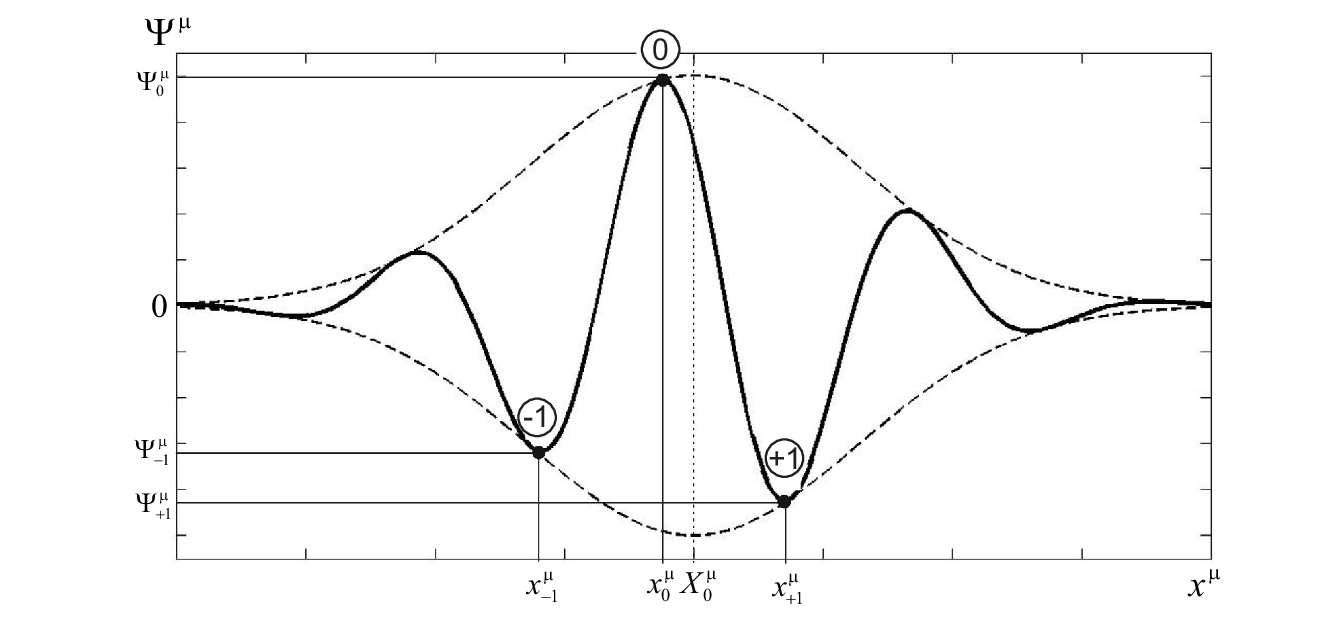}
\caption{Schematic view of the wave profile within the packet. The dashed lines show the envelope curves corresponding to the positive and negative values of the envelope function. The symbols $x_{0}^{\mu } $ and $x_{\pm 1}^{\mu } $ label the positions of the main peak and its two adjacent troughs, respectively. The position $x_{0}^{\mu } $ of the global half-wave does not coincide with the center of the envelope$X_{0}^{\mu } $ at a non-zero value of the initial phase $\delta ^{\mu } $ (see below)}.
\end{figure}

Differentiating function (16) and setting the result to zero, we obtain the transcendental equation
\begin{equation} \label{17}
\tan(k_{0} x^{\mu } +\delta ^{\mu } )=-\varepsilon k_{0} x^{\mu },
\end{equation}
which determines the positions of its extrema in the presence of the phase $\delta ^{\mu } $. Here,
\begin{equation} \label{18}
\varepsilon =\left(\frac{\sigma }{k_{0} } \right)^{2} <\left(\frac{\Delta k}{k_{0} } \right)^{2} <<1.
\end{equation}

 We will seek the solutions to Eq. (17) using the method of successive approximations with respect to the small parameter $\varepsilon $, but restrict ourselves to the linear approximation only. This yields
\begin{subequations}
\label{eq:main}
\begin{align}
    \label{eq:a}
     &k_{0} x_{-1}^{\mu } =-\left(1-\varepsilon \right)(\delta ^{\mu } +\pi ), \\
    \label{eq:b}
    &k_{0} x_{0}^{\mu } =-\left(1-\varepsilon \right)\delta ^{\mu }, \\
    \label{eq:c}
    &k_{0} x_{+1}^{\mu } =-\left(1-\varepsilon \right)(\delta ^{\mu } -\pi ).
\end{align}
\end{subequations}
Using (19), we find the interval width between the two side troughs:
\begin{equation} \label{20}
\Delta x^{\mu } \equiv x_{+1}^{\mu } -x_{-1}^{\mu } =2\pi (1-\varepsilon )/k_{0}.
\end{equation}
Since the right-hand side of this equation contains only the spectral parameters of the wavelet, the width of this interval must also be the same across all wave field components, and we can set
\begin{equation} \label{21}
\Delta x^{\mu } =\Delta x.
\end{equation}

Then, inverting (20), we conclude that the longitudinal wavenumber $k_{0} $ of the wavelet is related to its spectral parameter $\varepsilon $:
\begin{equation} \label{22}
k_{0} =\frac{2\pi }{\Delta x} (1-\varepsilon ).
\end{equation}

Interestingly, the wavelength $\lambda (\equiv 2\pi /k_{0} )$, which here represents the quasi-periodicity in the satellite's longitudinal direction, is no longer equal to the distance $\Delta x$ between adjacent troughs in this direction, but slightly exceeds this value.

Let us now specify the position of the main peak relative to the two side troughs. Based on Eqs. (19a)-(19c), we calculate the difference between the two intervals $\left(x_{1}^{\mu } -x_{0}^{\mu } \right)-\left(x_{0}^{\mu } -x_{-1}^{\mu } \right)$ and see that it vanishes. This implies that, to the linear approximation in $\varepsilon $, the main peak is located strictly midway between the side troughs:
\begin{equation} \label{23}
x_{+1}^{\mu } -x_{0}^{\mu } =x_{0}^{\mu } -x_{-1}^{\mu } =\frac{\Delta x}{2}.
\end{equation}

And finally, combining (19b) with (22), we obtain the following expression for the center of the packet envelope:
\begin{equation} \label{24}
X_{_{0} }^{\mu } =x_{_{0} }^{\mu } +\frac{\delta ^{\mu } \Delta x}{2\pi }.
\end{equation}

Thus, indeed, at a non-zero value of the phase $\delta ^{\mu } $, the center $X_{_{0} }^{\mu } $ of the wavelet envelope is shifted relative to the position $x_{_{0} }^{\mu } $ of the main half-wave.

Now, let us proceed to the analysis of the main oscillation amplitudes and substitute their positions (19a)-(19c) into the initial dependence (16). In the linear approximation with respect to $\varepsilon $, we obtain $\cos \left[k_{0} \left(x_{n}^{\mu } -X_{_{0} }^{\mu } \right)+\delta ^{\mu } \right]=(-1)^{n} $, $n=0,\pm 1$; therefore, we can get finally
\begin{subequations}
\label{eq:main}
\begin{align}
    \label{eq:sub_a}
     &\Psi _{-1}^{\mu } =-C^{\mu } \exp \left[-\frac{\varepsilon (\pi +\delta ^{\mu } )^{2} }{2} \right], \\
    \label{eq:sub_b}
    &\Psi _{0}^{\mu } =C^{\mu } \exp \left[-\frac{\varepsilon (\delta ^{\mu } )^{2} }{2} \right], \\
    \label{eq:sub_c}
    &\Psi _{+1}^{\mu } =-C^{\mu } \exp \left[-\frac{\varepsilon (\pi -\delta ^{\mu } )^{2} }{2} \right].
\end{align}
\end{subequations}
By multiplying and dividing Eqs. (25a) and (25c), and incorporating Eq. (25b), we finally yield
\begin{equation} \label{26}
\varepsilon =\frac{1}{\pi ^{2} } \ln \left[\frac{\left(\psi _{0}^{\mu } \right)^{2} }{\psi _{-1}^{\mu } \psi _{+1}^{\mu } } \right],      \end{equation}
\begin{equation} \label{27}
\delta ^{\mu } =\frac{\pi }{2} \, \frac{\ln \left(\frac{\psi _{+1}^{\mu } }{\psi _{-1}^{\mu } } \right)}{\ln \left[\frac{\left(\psi _{0}^{\mu } \right)^{2} }{\psi _{-1}^{\mu } \psi _{+1}^{\mu } } \right]}.
\end{equation}

Then, by inverting Eq. (25b), we can easily find the remaining unknown packet amplitude:
\begin{equation} \label{28}
C^{\mu } =\Psi _{0}^{\mu } \exp \left[\frac{\varepsilon (\delta ^{\mu } )^{2} }{2} \right].
\end{equation}

Thus, all five  wavelet parameters can be determined with help of three dominant half-waves. Specifically, the are the wavelength projection on the satellite orbit, the amplitude and the phase of the wavelet, the spatial center of the envelope, as well as the spectral parameter $\varepsilon $. They are sufficient to fully reconstruct the form of expression (16):
\begin{equation} \label{29}
\Psi ^{\mu } \left(x^{\mu } \right)=C^{\mu } \exp \left[-\varepsilon \frac{(kx^{\mu } )^{2} }{2} \right]\cos (k_{0} x^{\mu } +\delta ^{\mu } ),\quad x^{\mu } =X-X_{0}^{\mu }.
\end{equation}

Let us now analyze in more detail the parameters $\varepsilon $ and $\delta ^{\mu } $ that enter into Eq. (29). To do this, we use their definitions given in (26) and (27).

The parameter $\varepsilon $ is always positive and reaches its minimum, zero value for purely monochromatic wave where all amplitudes are equal, in particular, $\Psi _{0}^{\mu } =-\Psi _{\pm 1}^{\mu } $. This spectral parameter  acts as a modulation factor that describes the decay depth of the side half-waves relative to the amplitude of the main one.

The phase parameter $\delta ^{\mu } $ vanishes if the amplitudes of the side half-waves are equal in the magnitude. We will refer to this case as symmetric. Conversely, if the amplitude of one of the side half-waves approaches the amplitude of the central one, an antisymmetric profile emerges. Visually, this case is identified by the presence of two identical, adjacent half-waves of opposite signs. In this scenario, the wavelet center is located strictly midway between these two half-waves, and the corresponding local phase $\delta ^{\mu } $ equals $\pm \pi /2$.

Thus, the wavelet phase $\delta ^{\mu } $ indeed reflects the internal symmetry of the dominant half-waves within the wavelet and serves as a measure of this symmetry.

\section{Simulation results utilizing theoretical wavelet parameters}

In the previous section, we established that the wavelet parameters can be uniquely reconstructed from the data of its three dominant half-waves. The latter, in turn, can be determined directly from observational data. To evaluate the agreement between theory and experiment, as well as to verify the reliability of the theoretical results, we return here to the experimental profiles from Fig. 1. By applying the formulas derived above, we can now determine the internal wavelet parameters within the four local spatial domains. Based on these parameters, we reconstruct the individual wave profiles for each component and subsequently sum them to obtain the final synthesized result.

 In Figure 4, all experimental profiles are shown as solid lines in comparison with the theoretical calculations, represented by dashed lines. The dotted lines show the envelope curves corresponding to the positive and negative values of the envelope function within each wavelet. As can be seen, theory and experiment agree very well within the intervals of the second and fourth wavelets. In the region of the third wavelet, the discrepancies are minor, whereas for the first wavelet, they become significant. This indicates that the dominant half-waves for the first region were incorrectly identified due to the influence of an adjacent lower-amplitude packet. Consequently, the data within the initial interval are not suitable for a reliable analysis.
\begin{figure}[htbp]
\centering
\includegraphics[width=1.0\columnwidth]{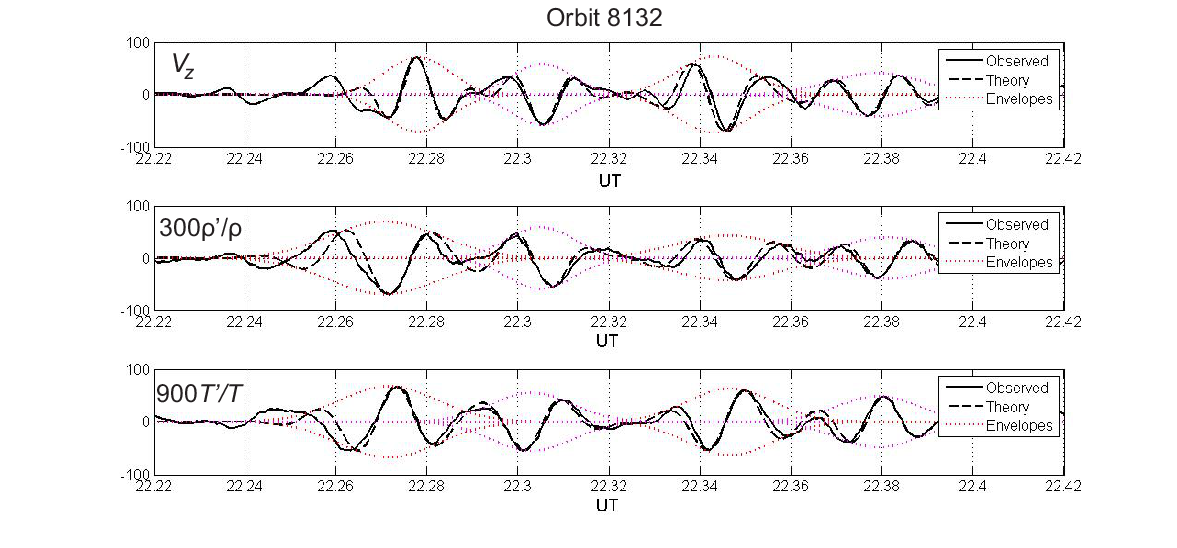}
\caption{Experimental and theoretical profiles of the physical quantities $V_{z} $, $\rho '/\rho $ and $T'/T$ for orbit 8132 of DE2 satellite (southern polar cap crossing).}
\end{figure}

Figure 5 presents the observational data for the same physical characteristics recorded during two additional orbits of the same satellite over the Earth's polar caps. (Data from this satellite are available via \url{ftp://nssdcftp.gsfc.nasa.gov/spacecraft\_data}). A common feature of these orbits is that the $\rho '/\rho $ and $V_{z} $ profiles exhibit nearly in-phase behavior (see the upper panel), while a phase shift of $\pi /2$ is almost always observed between the $T'/T$ and $V_{z} $ profiles (see the bottom panel of this Figure). The latter is highlighted by vertical dotted lines, demonstrating that the extrema of the $T'/T$ curve consistently align with the zero-crossings of the $V_{z} $ curve.
\begin{figure}[htbp]
\centering
\includegraphics[width=1.0\columnwidth]{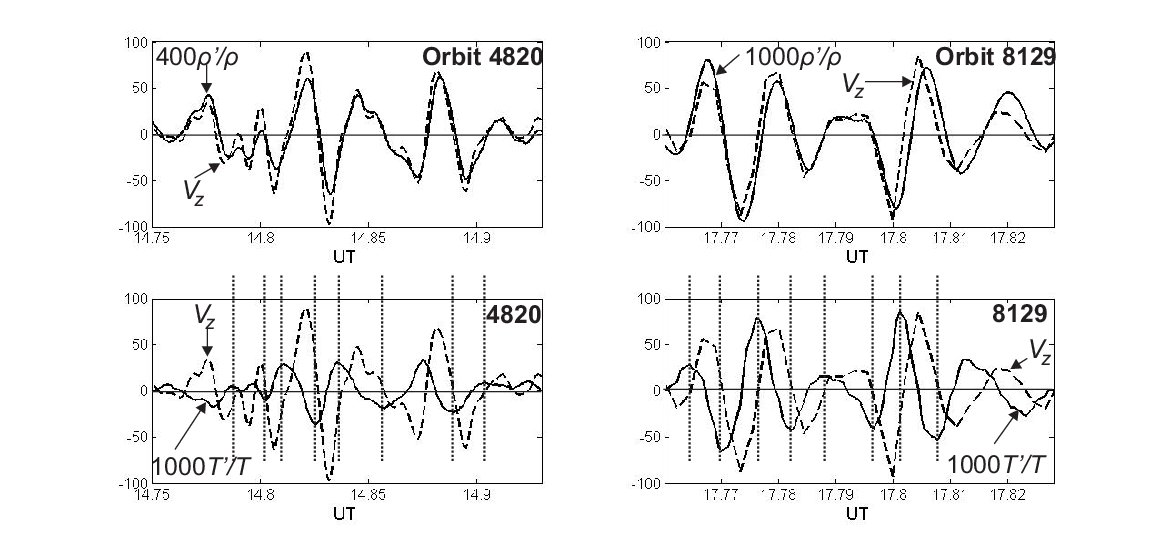}
\caption{Wave profiles of $\rho '/\rho$, $T'/T$, and $V_{z}$ for orbits 4820 and 8129 of the low-Earth orbit DE2 satellite, corresponding to its crossings of the northern and southern polar caps, respectively. The vertical dotted lines indicate the positions of the $T'/T$ extrema, which nearly coincide with the zero-crossings of the $V_{z} $ function.}
\end{figure}

For these two orbits, the procedure for generating the theoretical curves to subsequently compare them with the observational data is completely analogous to the method described above. The results are shown in Figure 6, where a good agreement between theory and experiment is also observed.
\begin{figure}[htbp]
\centering
\includegraphics[width=1.0\columnwidth]{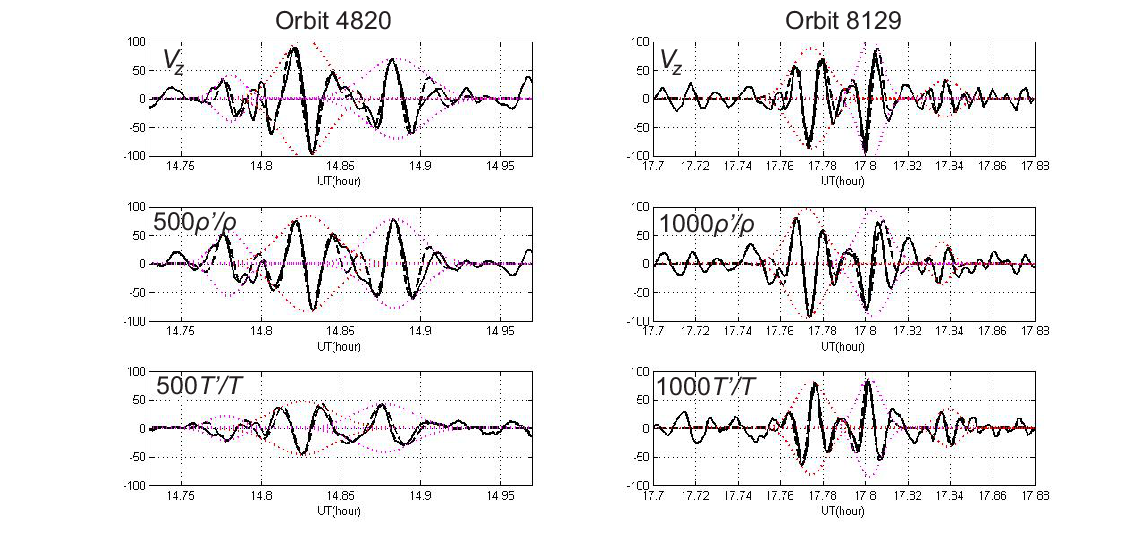}
\caption{Experimental and theoretical profiles of $V_{z} $, $\rho '/\rho $, and $T'/T$ for orbits 4820 (left panels) and 8129 (right panels), represented by solid and dashed lines, respectively. The dotted curves correspond to the wavelet envelopes.}
\end{figure}

All the comparative plots in this section indicate that the proposed approach, which utilizes wave patterns as fundamental building blocks, possesses high physical fidelity. The successful reproduction of the fine structure of disturbances across different orbits verifies the reliability of the model and  its ability to accurately describe the spatial evolution of wave processes in the polar ionosphere.

\section{Phase shifts in wave-packet representation}

The approach described above allows for reconstructing the wave profile of each measured component independently. Next, we  determine the phase shift between these components. To do this, we must return from the symmetry phase $\delta ^{\mu } $ to the original initial phase $\varphi ^{\mu } $ using the relationship given in Eq. (25). It should be emphasized that the carrier wavenumber $k_0$ entering this expression must be identical for all measured wave field components; otherwise, the physical meaning of the inter-component phase shift is lost. As a result, for the phase difference between the wave field components $\mu'$ and $\mu$, we obtain:

$$\Delta _{\mu ',\mu } \equiv \varphi ^{\mu '} -\varphi ^{\mu } =\delta ^{\mu '} -\delta ^{\mu } +k_0\left(X_{0}^{\mu } -X_{0}^{\mu '} \right).$$

In the following, the wavenumber $k_0$ is determined from the component whose observed profile is most clearly pronounced (for other components, these values are reasonably close, although they do not coincide identically). In the DE2 satellite dataset, the velocity and temperature were measured by the same experiment (WATS), whereas the density profile was obtained by a different experiment (NACS). Because these two experiments operated with different sampling characteristics, we selected the relative temperature fluctuations $T'/T$ as the reference component, given that this profile proved to be the smoothest. Similarly, another spectral parameter, the weighting factor $\varepsilon$ considered below, is also chosen to be unified for all components.

Now, we substitute Eqs. (22) and (24) into this last expression. Subsequently, we apply relations (20) and (21) to finally obtain
\begin{equation} \label{30}
\Delta _{\mu ',\mu } =\varepsilon \left(\delta ^{\mu '} -\delta ^{\mu } \right)+\left(1-\varepsilon \right)\frac{2\pi \left(X_{0}^{\mu } -X_{0}^{\mu '} \right)}{X_{+1}^{\mu '} -X_{-1}^{\mu '} }.
\end{equation}

 Formula (30) demonstrates that the phase shift between the profiles of the two wave field components is a weighted function of two factors, governed by the small weighting term $\varepsilon$. The first factor compares the internal symmetry-related phases of the corresponding components and enters with a small weight. The second, dominant factor represents the classical phase difference between the respective central extrema of two waves having the same periods. While the classical factor dominates, the symmetry-related correction is essential for high-precision phase measurements.

 To estimate the possible deviation of the phase difference (30) from its classical counterpart
 \begin{equation*}
 \Delta^{(\rm cl)}_{\mu ',\mu } =\frac{2\pi \left(X_{0}^{\mu } -X_{0}^{\mu '}\right)}{X_{+1}^{\mu '} -X_{-1}^{\mu '}},
 \end{equation*}
 it is convenient to rewrite Eq. (30) as
\begin{equation*}
 \Delta _{\mu ',\mu }-
\Delta^{(\rm cl)}
_{\mu ',\mu }=\varepsilon \left(\delta ^{\mu '} -\delta ^{\mu } -\Delta^{(\rm cl)}
_{\mu ',\mu } \right).
\end{equation*}
Theoretically, this deviation reaches its maximum when all phases on the right-hand side of this expression attain their limiting values ($\pm\pi/2$, $\pm\pi/2$, and $\pm\pi$, respectively) and their signs are chosen to maximize the absolute value of the right-hand side. Therefore, the theoretical maximum possible deviation cannot exceed $2\pi\varepsilon$.

As an example, Table 1 presents the true phase shift values $\Delta_{T,V_{z}}$ ($\Delta_{\rho,V_{z}}$) between the $T'/T$ ($\rho'/\rho$) and $V_{z}$ profiles for all four wave packets of orbit 8132 in comparison with the classical phase shift given in parentheses. The values of the parameter $\varepsilon$ for all four packets are calculated using Eq. (26) and are approximately equal to 0.07, 0.1, 0.09, and 0.07, respectively. For these values, the theoretical limit $2\pi\varepsilon$ corresponds to approximately $30^\circ$. Table 1 shows that the actual deviations observed on this orbit do not exceed $13^\circ$, which is fully consistent with the theoretical estimate. Importantly, even such moderate deviations remain significant for determining the true phase shift between the components.

\begin{table}[htbp]
  \centering
  \caption{Reconstructed phase shifts for the observed wave packets.}
  \label{tab:phase_shifts}
  \vspace{0.5em}
  \begin{tabular}{|c|c|c|}
    \hline
    Packet Number & $\Delta_{T, V_z}$ (deg) & $\Delta_{\rho, V_z}$ (deg) \\ \hline
    1 & 79 (90) & $-66$ ($-56$) \\ \hline
    2 & 86 (90) & $-44$ ($-45$) \\ \hline
    3 & 103 (90) & $-32$ ($-44$) \\ \hline
    4 & 87 (90) & $-50$ ($-52$) \\ \hline
  \end{tabular}
\end{table}

The values of the parameter $k_0$ for these four wave packets are not presented here because they do not hold any physical meaning: the profiles measured by the onboard satellite instruments and used to calculate $k_0$ in our consideration are defined in terms of universal time  rather than a spatial coordinate, as theoretically required. However, this circumstance has no effect on the method for determining the phase shift between the wave field components, as the latter remains invariant under the choice of coordinate system (whether spatial or temporal).

The  Table demonstrates that the density profile $T'/T$ lags behind the profile $V_{z} $ in all wave packets by a phase angle close to 90 degrees. A similar phase difference was visually observed for the other two previously analyzed orbits of the same satellite (see Figure 6). As will be shown below, this specific shift close to 90 degrees plays a central role in determining the full spectrum of the disturbances over Earth's polar caps.

\section{Reconstruction of AGW spectral parameters from phase shifts in observed wave field components}

Within the wave-packet framework, the spectral structure of a wave packet is primarily governed by its carrier wave. Since this carrier wave is treated as monochromatic, all fundamental relations derived in Sec. 2 are fully applicable to it. In our study, Hines' model serves as the background theoretical framework. Within this approach, knowing the two measured phase shifts $\Delta_{\rho,V_z}$ and $\Delta_{T,V_z}$ is entirely sufficient to determine the remaining spectral parameters of the carrier wave, including the intrinsic frequency, wave vector components, and propagation direction (Klymenko et al., 2021).  Crucially, as shown in Sec. 4, these phase shifts are invariant under the transformation between spacecraft-centered and Earth-fixed coordinate systems.

The results presented in this Section partially parallel those from our work cited above, and we only adapt them to the specific problem under consideration.

Indeed, substituting these phase constants into Eqs. (7a) and (7b) yields a linear inhomogeneous system of two equations with respect to two dimensionless unknowns, $k_{z} H$ and $(c_{s} k_{x} /\omega )^{2} $. Solving this system, we obtain
\begin{equation} \label{31}
k_{z} H=\frac{1}{2} _{} \frac{\gamma -2}{\left(\gamma -1\right)_{} \tan\Delta _{T,V_{z} } +\tan\Delta _{\rho ,V_{z} } },
\end{equation}
\begin{equation} \label{32}
\left(\frac{c_{s} k_{x} }{\omega } \right)^{2} =\frac{\gamma }{2} _{} \frac{\tan\Delta _{T,V_{z} } +\tan\Delta _{\rho ,V_{z} } }{\left(\gamma -1\right)_{} \tan\Delta _{T,V_{z} } +\tan\Delta _{\rho ,V_{z} } }.     \end{equation}
These two relations are insufficient to determine three spectral parameters $k_{z} $, $k_{x} $, and $\omega $; hence, we employ the dispersion relation (3), rewritten in the following equivalent form:
\[\left(k_{x} H\right)^{2} \left[\left(\frac{\omega ^{} }{c_{s} k_{x} } \right)^{2} -1\right]=\left[\frac{1}{4} -\frac{\gamma -1}{\gamma ^{2} } _{} \left(\frac{c_{s} k_{x} }{\omega } \right)^{2} \right]+\left(k_{z} H\right)^{2} .\]
Substituting (31) and (32) into this relation yields
\begin{equation} \label{33}
\left(k_{x} H\right)^{2} =\frac{\tan\Delta _{T,V_{z} } +\tan\Delta _{\rho ,V_{z} } }{\tan\Delta _{T,V_{z} } -\tan\Delta _{\rho ,V_{z} } } \times \frac{(\gamma -1)^{2} \left(1+\tan^{2} \Delta _{T,V_{z} } \right)-\left(1+\tan^{2} \Delta _{\rho ,V_{z} } \right)}{4\left[\left(\gamma -1\right)\tan\Delta _{T,V_{z} } +\tan\Delta _{\rho ,V_{z} } \right]^{2} }.
\end{equation}
The remaining unknown angular frequency $\omega $ can then be found by solving Eqs. (32) and (33):
\begin{equation} \label{34}
\left(\frac{\omega }{N} \right)^{2} =\frac{\gamma }{2(\gamma -1)} _{} \frac{(\gamma -1)^{2} \left(1+\tan^{2} \Delta _{T,V_{z} } \right)-\left(1+\tan^{2} \Delta _{\rho ,V_{z} } \right)}{\left(\tan\Delta _{T,V_{z} } -\tan\Delta _{\rho ,V_{z} } \right)\left(\left(\gamma -1\right)\tan\Delta _{T,V_{z} } +\tan\Delta _{\rho ,V_{z} } \right)}.
\end{equation}

Thus, all spectral parameters of the AGW are now determined; they are defined by equations (31), (33), and (34) for the known values of the angles $\Delta _{T,V_{z} } $ and $\Delta _{\rho ,V_{z} } $.

The wave propagation angle $\theta $ relative to the horizon, $\tan\theta =k_{z} /k_{x} $, follows from equations (31) and (33). Summing these equations yields the squared magnitude of the wave vector: $\left|\mathbf{k}\right|^{2} =k_{x}^{2} +k_{z}^{2} $. The squares of the horizontal ($\omega / k_{x}$) and vertical ($\omega / k_{z}$) phase velocities of the wave are given by:
\[\left(\frac{\omega }{k_{x} } \right)^{2} =\frac{2c_{s}^{2} }{\gamma } \frac{\left(\gamma -1\right)_{} \tan\Delta _{T,V_{z} } +\tan\Delta _{\rho ,V_{z} } }{\tan\Delta _{T,V_{z} } +\tan\Delta _{\rho ,V_{z} } } ,     \]

\[\left(\frac{\omega }{k_{z} } \right)^{2} =\frac{c_{s}^{2} }{\gamma (\gamma -2)}\frac{(\gamma -1)^{2} \left(1+\tan^{2} \Delta _{T,V_{z} } \right)-\left(1+\tan^{2} \Delta _{\rho ,V_{z} } \right)}{\tan\Delta _{T,V_{z} } -\tan\Delta _{\rho ,V_{z} } } . \]

Similarly, the expressions for the wave group velocity can be derived by differentiating the dispersion relation (3) and incorporating (31), (33), and (34).

Equations (32) and (33) also allow for a straightforward construction of the phase diagram, as their right-hand sides must be non-negative for freely propagating waves. It should also be taken into account that the acoustic region of the allowed spectrum lies above the Lamb curve $\omega=c_{s}k_x$, whereas the gravity wave region is located below this curve (Hines, 1960). Applied to Eq. (33), the latter implies that a value of unity on its right-hand side separates these two types of AGWs. Solving all the resulting inequalities geometrically leads to the diagram shown in Figure 7. For details, see (Klymenko et al., 2021).

\begin{figure}[htbp]
\centering
\includegraphics[width=1.0\columnwidth]{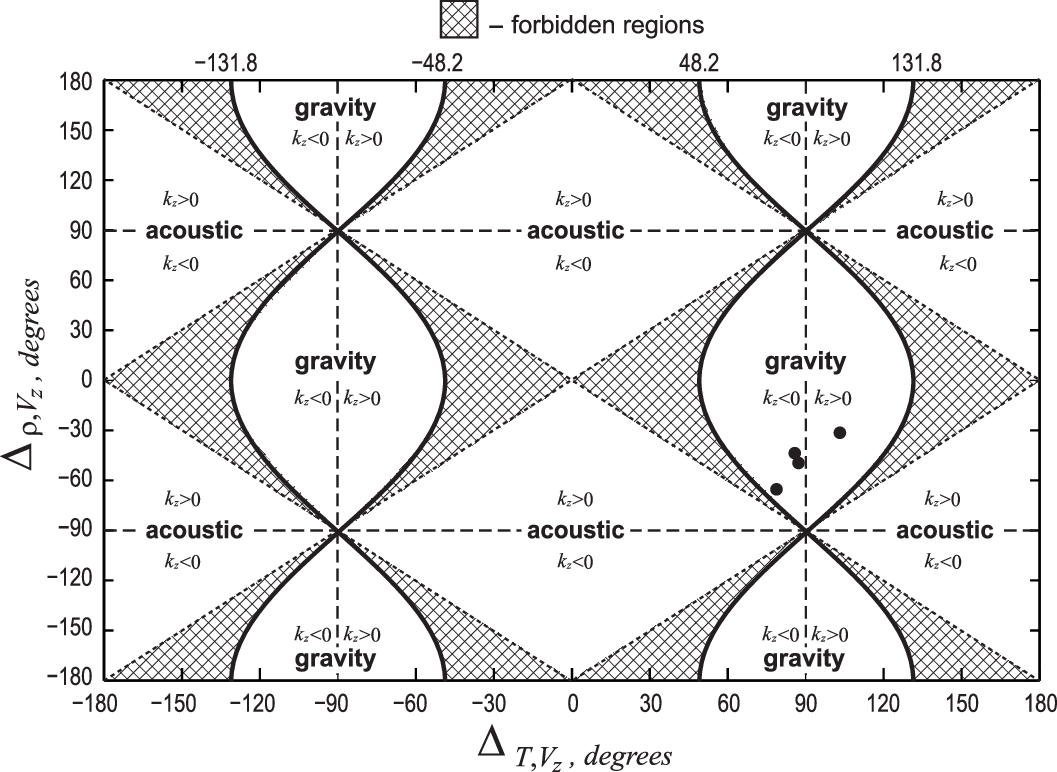}
\caption{Existence regions for different types of freely propagating AGWs on the phase plane $(\Delta _{\rho ,V_{z} },\Delta _{T,V_{z} } )$ for adiabatic index $\gamma=5/3$ . The forbidden regions are shaded. The additional markers (black circles) show phase points whose physical meaning becomes clear in the next section.}
\end{figure}

This diagram visually separates the regions of existence for the acoustic and gravity branches of AGWs from the forbidden region, making it highly usefull for analysis. After calculating the angles $\Delta _{T,V_{z} } $ and $\Delta _{\rho ,V_{z} } $ from observational data one simply needs to plot this point on the phase diagram. This way immediately identifies the wave type (acoustic or gravity) and determines the direction of the wave propagation relative to the horizon (upward or downward). The remaining spectral parameters are then reconstructed analytically using the formulas presented in this section.

\section{Spectral characteristics of the profiles}

Using the data from Table 1, we plot four phase points of orbit 8132 on the phase diagram. All four points fall into the lower half of the gravitational domain and correspond to the phase points previously marked by black circles in Fig. 7. The spectral data corresponding to these points are summarized in Table 2. Thus, in all cases we observe medium-scale gravity-type waves propagating at a moderate angle to the horizon.

\begin{table}[htbp]
  \centering
  \caption{Spectral parameters of waves for orbit 8132.}
  \label{tab:spectral_params_8132}
  \vspace{0.5em}
  \begin{tabular}{|c|c|c|c|c|c|}
    \hline
    Packet Number & $k_z H$ & $k_x H$ & $|\mathbf{k}| H$ & $\omega / N$ & $\theta$ (deg) \\ \hline
    1 & $-0.14$ & $0.65$ & $0.66$ & $0.94$ & $-12$ \\ \hline
    2 & $-0.02$ & $0.52$ & $0.52$ & $0.92$ & $-2$  \\ \hline
    3 & $0.05$  & $0.44$ & $0.45$ & $0.84$ & $6$   \\ \hline
    4 & $-0.01$ & $0.52$ & $0.52$ & $0.92$ & $-5$  \\ \hline
  \end{tabular}
\end{table}

Let us find the spectral parameters for the wave packets observed on orbits 4820 and 8129, shown in Figure 6. We take advantage of the fact noted earlier that the profiles $\rho '/\rho $ and $V_{z} $ are nearly synchronous, while the phase shift between $T'/T$ and $V_{z} $ is nearly equal to $\pi /2$. By applying the phase diagram, we can see that these cases correspond to the points falling near the geometric center of the same gravitational part on the diagram as before.

For the profile configurations mentioned above, the key factor is that the value of $\tan\Delta _{T,V_{z} } (>>1)$ dominates over the others. By considering only this term in the formulas above, one can easily obtain the following linear estimations for the spectral wave parameters at $\Delta_{T,V_{z}}\rightarrow \pi/2$:
\begin{equation*}
k_{z} H=\frac{\gamma -2}{2\left(\gamma -1\right)} \left(\frac{\pi }{2} -\Delta _{T,V_{z} } \right), \quad k_{x} H=\frac{1}{2},\quad  \left|\mathbf{k}\right|H= \frac{1}{2} ,
\end{equation*}
\begin{equation*}
\frac{\omega }{N}=\sqrt{\frac{\gamma }{2} } \approx 0.91,\quad  \tan\theta =\frac{\gamma -2}{\gamma -1} \left(\frac{\pi }{2} -\Delta _{T,V_{z} } \right).
\end{equation*}
That is, we return to the case of the medium-scale waves ($\left|\vec{k}\right|H=0.5$) of the gravity-type ($\omega <N$) that propagate almost horizontally ($\theta \approx 0$).

It should be noted that the peculiar spectral properties of AGWs over polar caps were reported previously (Johnson et al., 1995; Innis and Conde, 2002; Fedorenko et al., 2025). In those studies, the AGW characteristics were also analyzed based on DE2 polar satellite measurements. To evaluate the wave properties, the authors employed theoretical framework within the monochromatic plane-wave approximation. Fedorenko et al. (2025) attributed the observed features and large AGW amplitudes in the polar region to wave spectrum filtering in inhomogeneous wind flows.

\section{Discussion and conclusions}

The data processing method proposed in this work is based on the approach that involves preliminarily partitioning the entire spatial disturbance of the wave field components into separate segments. These segments can be interpreted as physical patterns of individual wave events. The method operates exclusively with the geometric characteristics of these local regions, namely the central extremum and two adjacent out-of-phase oscillations of smaller amplitude. The criterion for identifying these short isolated segments as a wave packet is the equidistance of the two adjacent side extrema of opposite sign from the central maximum (or minimum). This property can be visually verified directly from the observational data.

If a given local segment is identified as an individual wave packet, the wave packet theory is applied to it. This theory is physically transparent, avoids mathematical complexities such as the inverse Fourier transform, and yields the classical expression for a quasi-monochromatic wave packet modulated by a slowly varying envelope.

Next, we utilized the slowly varying nature of the envelope, approximating it with a Gaussian function -- a standard choice for the spectral-temporal analysis of wave disturbances. This yields an analytical expression for the wave profile geometry within the packet, making it possible to identify all five of its primary parameters solely from the positions and amplitudes of the three aforementioned dominant half-waves. Alternatively, this Gaussian approximation can be replaced by any other explicit envelope shape featuring a specific localization center and a scale parameter chosen based on a priori physical considerations. Such a wavelet configuration is characterized by the same set of five packet parameters as our proposed Gaussian profile. These parameters can likewise be uniquely determined from its dominant peaks. In other words, the proposed geometric approach for reconstructing packet parameters from three dominant half-waves is universal, differing only in the specific analytical expressions used for this reconstruction.

Regardless of the envelope model, two of the five packet parameters are spectral (specifically the carrier wavenumber and the spectral bandwidth parameter) and are common to all observed components of the wave field. In contrast, the remaining three parameters characterize the local geometric shape of the packet (its spatial center, amplitude, and initial phase) and are determined individually for each component. Reconstructing the packet parameters using the formulas derived makes it possible to fully restore the analytical form of the wave profile. When plotted numerically, this reconstructed profile virtually coincides with the experimental profiles obtained directly from satellite observations.

Based on the approach proposed, we succeed in establishing the true physical phase shift between different atmospheric wave field components (such as fluctuations in temperature, density, or wave velocity components). This true phase shift proves to be a weighted combination of two factors. The first factor relates to the difference between the phases that account for the geometric asymmetry of the packets, whereas the second factor is the classical phase shift, which is traditionally used to determine the phase difference between monochromatic waves of the same period.

Although the second factor contributes the dominant share, the first (symmetry-related) factor provides a moderate correction. This effectively implies that the true phase shift between the components differs slightly from what is visually apparent in the observed wave profiles. In other words, the symmetry-related factor is of critical importance for the accurate analysis of disturbance properties, provided that this analysis is based on utilizing phase shifts between different components of the wave field.

This work utilizes precisely such an approach, which makes it possible to determine the spectral parameters of a disturbance observed in an ideally stratified atmosphere from the phase shifts between density (or temperature) fluctuations and the vertical velocity component. This approach was applied to observational data gathered over several orbits of the low-Earth-orbit satellite DE2 during its transits over the Earth's polar caps. The obtained results demonstrated that the disturbances recorded by the satellite correspond to mid-latitude gravity-type AGWs propagating almost horizontally.

Although similar conclusions regarding the spectral characteristics of AGWs over polar regions have been previously documented in the literature, the proposed method possesses a significantly broader scope of application. It can be effectively utilized for signal processing in any spatial (or temporal) domains that can be identified as separate, isolated wave packets. When applied directly to a single measured disturbance, this method yields the spectral components of the profile along the satellite orbit. Furthermore, if applied to different wave field components, it enables the calculation of the phase shifts between them. Ultimately, if there exists a theoretical framework that couples the spectral parameters of the physical process to phase shifts between some observed physical components -- as is the case here for the model of an ideally stratified atmosphere -- one can fully reconstruct the complete spectra of these waves.

The practical application of our method to the analysis of other wave processes will be detailed in our subsequent publications.

\section*{Acknowledgments}

This work was supported by the State Institution National Antarctic Scientific Center within the framework of the State Special-Purpose Research Program in Antarctica for 2011--2027.

\section*{Data availability}

The satellite measurement data supporting the findings of this study are publicly available via \url{ftp://nssdcftp.gsfc.nasa.gov/spacecraft_data} as specified in Section 7.


\begin{thebibliography}{20}

\bibitem[Achatz et al.(2024)]{Achatz2024}
Achatz, U., Alexander, M.J., Becker, E., Chun, H.-Y., D\"{o}rnbrack, A., Holt, L., Plougonven, R., Polichtchouk, I., Sato, K., Sheshadri, A., Stephan, C.C., van Niekerk, A., Wright, C.J., 2024. Atmospheric Gravity Waves: Processes and Parameterization. J. Atmos. Sci. 81 (2), 237--262. https://doi.org/10.1175/JAS-D-23-0210.1.

\bibitem[Afraimovich et al.(2010)]{Afraimovich2010}
Afraimovich, E.L., Edemsky, I.K., Voeykov, S.V., Yasukevich, Yu.V., Zhivetiev, I.V., 2010. MHD nature of ionospheric wave packets generated by the solar terminator. Geomagn. Aeron. 50 (1), 79--95.

\bibitem[Cheremnykh et al.(2023)]{Cheremnykh2023}
Cheremnykh, O., Fedorenko, A., Voitsekhovska, A., Selivanov, Y., Ballai, I., Verth, G., Fedun, V., 2023. Atmospheric waves disturbances from the solar terminator according to the VLF radio stations data. Adv. Space Res. 72 (11), 4825--4835. https://doi.org/10.1016/j.asr.2023.08.036.

\bibitem[Cheremnykh et al.(2024)]{Cheremnykh2024}
Cheremnykh, O.K., Fedorenko, A.K., Kryuchkov, E.I., Klymenko, Y.O., Zhuk, I.T., 2024. Developing the Models of Acoustic-Gravity Waves in the Upper Atmosphere (Review). Kinemat. Phys. Celest. Bodies 40 (1), 1--14. https://doi.org/10.15407/kfnt2024.01.003.

\bibitem[Fedorenko and Kryuchkov(2014)]{Fedorenko2014}
Fedorenko, A.K., Kryuchkov, E.I., 2014. Observed features of acoustic gravity waves in the heterosphere. Geomagn. Aeron. 54, 109--116. https://doi.org/10.1134/S0016793214010022.

\bibitem[Fedorenko et al. (2025)]{Fedorenko2025}
Fedorenko, A., Kryuchkov, E., Voitsekhovska A., 2025.  Wind as an energy source of polar atmospheric gravity waves. Ukrainian Antarctic Journal, 23(1), 19--31. https://doi.org/10.33275/1727--7485.1.2025.741.

\bibitem[He et al.(2024)]{He2024}
He, M., Chau, J.L., Stober, G., Hall, C.M., 2024. Limitations of wavelet analysis in retrieving atmospheric wave profiles from satellite and radio occultation data. Atmos. Meas. Tech. 17 (15), 4659--4672.

\bibitem[Hines(1960)]{Hines1960}
Hines, C.O., 1960. Internal atmospheric gravity waves at ionospheric heights. Can. J. Phys. 38, 1441--1481. https://doi.org/10.1139/p60-150.

\bibitem[Innis and Conde(2002)]{Innis2002}
Innis, J.L., Conde, M., 2002. Characterization of acoustic-gravity waves in the upper thermosphere using Dynamics Explorer 2 Wind and Temperature Spectrometer (WATS) and Neutral Atmosphere Composition Spectrometer (NACS) data. J. Geophys. Res. Space Phys. 107 (A12). https://doi.org/10.1029/2002JA009370.

\bibitem[Johnson et al.(1995)]{Johnson1995}
Johnson, F.S., Hanson, W.B., Hodges, R.R., Coley, W.R., Carignan, G.R., Spencer, N.W., 1995. Gravity waves near 300 km over the polar caps. J. Geophys. Res. 100, 23993--24002.

\bibitem[Klymenko et al.(2021)]{Klymenko2021}
Klymenko, Yu.O., Fedorenko, A.K., Kryuchkov, E.I., Cheremnykh, O.K., Voitsekhovska, A.D., Zhuk, I.T., 2021. Identification of Acoustic-Gravity Waves According to the Satellite Measurement Data. Kinemat. Phys. Celest. Bodies 37 (6), 273--283. https://doi.org/10.3103/s0884591321060052.

\bibitem[Li et al.(2022)]{Li2022}
Li, X., Liu, A.X., Lu, X., 2022. A review of gravity wave extraction and analysis techniques in atmospheric observations. Atmos. Res. 265, 105912.

\bibitem[Liu and Lauritzen(2024)]{Liu2024}
Liu, H.-L., Lauritzen, P.H., 2024. Impacts of Gravity Waves on the Thermospheric Circulation and Composition. Geophys. Res. Lett. e2023GL107453. https://doi.org/10.1029/2023GL107453.

\bibitem[Medvedev and Yi\u{g}it(2019)]{Medvedev2019}
Medvedev, A.S., Yi\u{g}it, E., 2019. Gravity Waves in Planetary Atmospheres: Their Effects and Parameterization in Global Circulation Models. Atmosphere 10 (9), 531. https://doi.org/10.3390/atmos10090531.

\bibitem[Nappo(2002)]{Nappo2002}
Nappo, C.J., 2002. An introduction to atmospheric gravity waves, 2nd ed. Elsevier Science.

\bibitem[Sakib and Yi\u{g}it(2022)]{Sakib2022}
Sakib, M.N., Yi\u{g}it, E., 2022. A brief overview of gravity wave retrieval techniques from observations. Front. Astron. Space Sci. 9, 824875.

\bibitem[\v{S}auli et al.(2007)]{Sauli2007}
\v{S}auli, P., Roux, S.G., Abry, P., Bo\v{s}ka, J., 2007. Acoustic--gravity waves during solar eclipses: Detection and characterization using wavelet transforms. J. Atmos. Sol.-Terr. Phys. 69 (18), 2465--2484.

\bibitem[Schoennagel et al.(2006)]{Schoennagel2006}
Schoennagel, A., Lu, G., Kuo, C., 2006. Atmospheric response and bow-wave packet generation during solar eclipses. J. Atmos. Sol.-Terr. Phys. 68 (3-5), 586--596.

\bibitem[Vadas et al.(2023)]{Vadas2023}
Vadas, S.L., Becker, E., Figueiredo, C., Bossert, K., Harding, B.J., Gasque, L.C., 2023. Primary and secondary gravity waves and large-scale wind changes generated by the Tonga volcanic eruption on 15 January 2022: Modeling and comparison with ICON-MIGHTI winds. J. Geophys. Res. Space Phys. 128 (2), e2022JA031138. https://doi.org/10.1029/2022JA031138.

\bibitem[Wright et al.(2024)]{Wright2024}
Wright, C.J., Hindley, N.P., Hoffmann, L., 2024. Wavelet transformation capabilities and limitations in non-stationary geophysical signal analysis. arXiv preprint arXiv:2404.18781.

\end{thebibliography}
\end{document}